\documentclass[10pt,conference]{IEEEtran}
\usepackage{soul}
\usepackage{hyperref}
\usepackage{multirow}
\usepackage{listings}
\usepackage{fancybox}
\usepackage{enumitem}
\usepackage{graphicx}
\usepackage{color}
\usepackage{array}
\usepackage{amsmath}
\usepackage{xspace}
\usepackage{url}
\usepackage{tikz}
\usepackage{caption}
\usepackage{pgfplots}
\usepackage{balance}
\usepackage{subcaption}
\pgfplotsset{compat=1.16}
\usepackage{pgf-pie}
\usetikzlibrary{pgfplots.statistics,calc}
\usepackage{algorithm}
\usepackage{algorithmic}
\usepackage{adjustbox}
\usepackage{makecell}

\usepackage{tcolorbox}
\makeatletter
\newcommand{\mybox}[1]{%
	\setbox0=\hbox{#1}%
	\setlength{\@tempdima}{\dimexpr\wd0+13pt}%
	\begin{tcolorbox}[boxrule=0.5pt, colback=white, arc=4pt,
		left=4pt,right=4pt,top=4pt,bottom=4pt,boxsep=0pt]
		#1
	\end{tcolorbox}
}

\definecolor{songcolor}{RGB}{191,191,191}

\IEEEoverridecommandlockouts
\usepackage{cite}
\usepackage{amsmath,amssymb,amsfonts}
\usepackage{graphicx}
\usepackage{textcomp}
\usepackage{xcolor}
\def\BibTeX{{\rm B\kern-.05em{\sc i\kern-.025em b}\kern-.08em
    T\kern-.1667em\lower.7ex\hbox{E}\kern-.125emX}}
\begin{document}

\title{Prompt Engineering or Fine-Tuning: \\ An Empirical Assessment of LLMs for Code}

\author{%
  \IEEEauthorblockN{%
    Jiho Shin\IEEEauthorrefmark{1},
    Clark Tang,
    Tahmineh Mohati\IEEEauthorrefmark{2},
    Maleknaz Nayebi\IEEEauthorrefmark{1},
    Song Wang\IEEEauthorrefmark{1},
    Hadi Hemmati\IEEEauthorrefmark{1}\IEEEauthorrefmark{2}
  }%
  \IEEEauthorblockA{\IEEEauthorrefmark{1} Lassonde School of Engineering, York University, Toronto, Canada}%
  \IEEEauthorblockA{\IEEEauthorrefmark{2} Schulich School of Engineering, University of Calgary, Calgary, Canada}%
  \IEEEauthorblockA{ \{jihoshin, mnayebi, wangsong, hemmati\}@yorku.ca, clark.f.tang@gmail.com, tahmineh.mohati1@ucalgary.ca}
}

\maketitle

\begin{abstract}
The rapid advancements in large language models (LLMs) have greatly expanded the potential for automated code-related tasks. Two primary methodologies are used in this domain: prompt engineering and fine-tuning. Prompt engineering involves applying different strategies to query LLMs, like \emph{ChatGPT}, while fine-tuning further adapts pre-trained models, such as \emph{CodeBERT}, by training them on task-specific data. Despite the growth in the area, there remains a lack of comprehensive comparative analysis between the approaches for code models.

In this paper, we evaluate \emph{GPT-4} using three prompt engineering strategies—basic prompting, in-context learning, and task-specific prompting—and compare it against 17 fine-tuned models across three code-related tasks: code summarization, generation, and translation. Our results indicate that \emph{GPT-4} with prompt engineering does not consistently outperform fine-tuned models. For instance, in code generation, \emph{GPT-4} is outperformed by fine-tuned models by 28.3\% points on the MBPP dataset. It also shows mixed results for code translation tasks.

Additionally, a user study was conducted involving 27 graduate students and 10 industry practitioners. The study revealed that \emph{GPT-4} with conversational prompts, incorporating human feedback during interaction, significantly improved performance compared to automated prompting. Participants often provided explicit instructions or added context during these interactions. These findings suggest that \emph{GPT-4} with conversational prompting holds significant promise for automated code-related tasks, whereas fully automated prompt engineering without human involvement still requires further investigation.

\end{abstract}


\section{Introduction}
\label{sec:intro}
Language models have gained significant interest, leading to numerous studies adapting their use for automated code-related tasks. Although originally developed for natural language, Large Language Models (LLMs) demonstrate potential in automated code-related tasks. Previous literature mainly focuses on pre-training on large corpora of Source Code (SC) and Natural Language (NL) and fine-tuning with task-specific datasets for various tasks. State-of-the-art LLMs, however, are trained using massive unsupervised corpora and use prompts for task querying. The large training corpus and numerous parameters enable LLMs to perform well on automated code-related tasks with simple prompts.

Prompting offers advantages over fine-tuning: it does not require labelled datasets, which are costly to acquire. Furthermore, running a prompt is less resource-intensive than fine-tuning an LLM. Despite its potential, prompt engineering is still in its early stages and lacks systematic studies on its performance compared to fine-tuned models in code tasks.

This paper presents a quantitative and qualitative investigation of \emph{OpenAI's ChatGPT}, specifically the latest version (\emph{GPT-4}), which has shown notable improvements over its predecessor, \emph{GPT-3.5}. We focus on three automated code tasks: code summarization (SC-to-NL), code generation (NL-to-SC), and code translation (SC-to-SC). These tasks are chosen for their commonality among developers and prevalence in the literature. We employ three automated prompting techniques (basic, in-context learning, and task-specific prompts) and compare them with 17 fine-tuned language models. Additionally, we surveyed 27 graduate students and 10 industry practitioners to gather qualitative insights through conversational prompts. 
We address the following research questions:

{\noindent \textbf{RQ1: How does \emph{GPT-4} with automated prompting compare to fine-tuned models in performance?}}
We quantitatively assess \emph{GPT-4} using three prompting strategies: (a) basic prompt, (b) in-context learning, and (c) task-specific prompt, comparing results with fine-tuned language models. 

\noindent \textbf{RQ2: How do participants perceive the usefulness of \emph{GPT-4} using a basic prompt?}
We ask participants to assess the basic prompting strategy qualitatively to understand its usefulness.

\noindent \textbf{RQ3: How do participants refine their prompts when interacting with \emph{GPT-4}?}
We investigate how participants refine prompts to achieve better results by analyzing their interaction logs and summarizing prompt evolution patterns.

\noindent\textbf{RQ4: What is the impact of different prompt evolution patterns?}
We investigate the impact of prompt evolution patterns derived from participant interactions.

Our quantitative analysis suggests that prompt-engineered \emph{GPT-4} does not consistently outperform fine-tuned LLMs in all tasks. For code summarization, \emph{GPT-4} with task-specific prompting outperforms the top fine-tuned model by 8.33\% points in the BLEU score. In code generation, \emph{GPT-4} outperforms fine-tuned models by 8.59\% points on the HumanEval dataset. However, it is outperformed by fine-tuned models by 28.3\% points on the MBPP dataset. In code translation, \emph{GPT-4} and fine-tuned models also show mixed results. Analysis of participant interactions reveals significant performance improvements when using conversational prompts compared to automated strategies, with improvements of 15.8\% points, 18.3\% points, and 16.1\% points for code summarization, generation, and translation, respectively. Participants frequently requested improvements, added context, or provided specific instructions to enhance \emph{GPT-4}'s output. Our results indicate that \emph{GPT-4} with conversational prompting holds potential for automated code-related tasks, but fully automated prompt engineering requires further development.

This paper contributes the following:
\begin{itemize}
\item The first empirical study comparing automated prompting strategies on \emph{GPT-4} with fine-tuned LLMs for three automated code-related tasks. 
\item A user study with 27 graduate students and ten industry practitioners exploring the evolution of conversational prompts in automated code-related tasks. 
\item Identification of gaps between conversational and automated prompts, and suggestions for leveraging LLMs in automated code-related tasks. 
\item Release of all study artifacts to facilitate replication and extension by other researchers\footnote{\url{https://github.com/shinjh0849/gpt4_ase_tasks}}. 
\end{itemize}

\section{Background and Related Work}
\label{sec:background}

\subsection{Fine-Tuning Language Models for Code}
\label{sec:lms}
The field of automating code-related tasks has increasingly adopted language models (LMs) \cite{wang2016automatically,white2016deep,yin2018tranx,liu2019deep,allamanis2018learning}. These methods offer significant advantages over traditional approaches such as domain-specific models, probabilistic grammars, and simple neural LMs. In recent years, LMs have been applied to various ASE applications, including code completion~\cite{wei2023copiloting}, code search \cite{gu2018deep}, code generation \cite{shin2021survey, shin2023good}, test case generation \cite{shin2023domain,shin2024retrieval}, test oracle generation \cite{shin2023assessing}, code summarization \cite{hu2018deep}, code translation \cite{sun2022code}, and automated program repair \cite{li2022dear}.

Pre-trained code models learn general-purpose code representations capturing lexical, syntactic, semantic, and structural information. Fine-tuning adapts these models to specific tasks using task-specific data, allowing them to outperform existing baselines. Numerous studies have leveraged pre-training on source code and natural language corpora, followed by fine-tuning for downstream tasks, e.g., code search, code summarization, code generation, etc. \cite{feng2020codebert,kanade2020learning,guo2022unixcoder,ahmad2021unified,ahmed2022multilingual}.

\subsection{Prompt Engineering in Software Engineering}
\label{sec:prompt-engineering}
Prompt engineering is an alternative to fine-tuning that adapts pre-trained LMs without requiring a supervised dataset. Instead, it uses prompts to treat different tasks as generation problems. These models, termed Large Language Models (LLMs), have larger corpora and more parameters. The advent of LLMs and prompt engineering has significantly improved task performance \cite{brown2020language,ouyang2022training,openai2023gpt4,touvron2023llama,touvron2023llama2,chowdhery2022palm,anil2023palm2}. Studies have explored LLMs and prompt engineering to tackle code tasks with various prompting strategies, such as basic prompting, in-context learning, task-specific prompting, chain-of-thought prompting, auto-prompting, and soft prompting \cite{khan2022automatic,gao2023constructing,wei2023copiloting,feng2023prompting,geng2024large,liu2023pre,liu2022makes,he2022hyperprompt,carta2023iterative,wei2022chain,shin2020autoprompt,hambardzumyan2021warp}.

Gao et al. \cite{gao2023constructing} investigated three key factors in in-context learning for code tasks: selection, order, and number of examples. They found that both similarity and diversity in example selection were crucial for performance and stability. Li et al. \cite{li2023nuances} studied \emph{ChatGPT}'s ability to find failure-inducing test cases, showing that performance improved drastically with correct guidance. Feng et al. \cite{feng2023prompting} introduced \emph{AdbGPT}, an LLM-based approach for reproducing bugs using few-shot learning and chain-of-thought reasoning. Kabir et al. \cite{kabir2023answers} analyzed \emph{ChatGPT}'s responses to Stack Overflow questions, finding over half of the answers incorrect and 77\% verbose. Geng et al. \cite{geng2024large} adopted in-context learning for code summarization, using customized strategies like selection and re-ranking to enhance performance.

Wang et al. \cite{wang2022no} compared prompt-tuning and fine-tuning on three code tasks (defect prediction, code summarization, and code translation) using two pre-trained models, \emph{CodeBERT} and \emph{CodeT5}. In contrast, we explored the effectiveness of \emph{GPT-4} using three prompting techniques (basic, in-context learning, and task-specific prompting) against 16 fine-tuned LLMs. The main difference between the studies lies in model size: they assessed models with 220M and 125M parameters, whereas we evaluated \emph{GPT-4} with 1.76T parameters. We also included a qualitative analysis surveying academia and industry participants to explore the impact of different prompting strategies.

Despite growing interest in prompt engineering, comparisons between fine-tuned and prompt-engineered LLMs in ASE tasks remain limited. Fine-tuning modifies pre-trained LM parameters to optimize a task, while prompt engineering crafts natural language queries to obtain the desired output without parameter changes. Fine-tuning can achieve better performance but requires more data and resources \cite{trad2024prompt,wang2022no}. Prompt engineering leverages LLMs' versatility but may face issues like inconsistency and insufficient domain knowledge optimization \cite{ouyang2023llm,trad2024prompt,li2023assisting}. To address this gap, we conducted the first quantitative and qualitative comparison of fine-tuning and prompt engineering methods.

\section{Empirical Study Setup}
\label{sec:emp_data}

\subsection{Downstream Tasks on Code Automation}
\label{sec:tasks}
We select three typical automated code-related software engineering tasks for the experiments, i.e., \textbf{Code summarization:} \cite{ahmed2022few, ahmed2024automatic, shi2022evaluation}The model generates a short natural language summary from a source code snippet  (SC-to-NL), \textbf{Code generation:} \cite{mu2024clarifygpt, li2022competition, dong2024self}The model generates the corresponding code snippet from a natural language description (NL-to-SC), and \textbf{Code translation:} \cite{luo2024bridging, pan2024lost, yang2024exploring} The model translates a source code snippet into another programming language (SC-to-SC).

These tasks assess \emph{GPT-4}'s ability to generate various modalities (SC and NL). The programming languages differ per task: code summarization involves \texttt{Ruby}, \texttt{JavaScript}, \texttt{Go}, \texttt{Python}, \texttt{Java}, and \texttt{PHP}; code generation targets \texttt{Python}; and code translation involves \texttt{Java} and \texttt{C\#}, covering both translation directions. The languages are chosen based on the benchmarks assessed, which are discussed in the following section.

\begin{table}[t!]
\centering
\caption{Stats of dataset used in quantitative analysis}
\label{tab:dataset}
\begin{adjustbox}{width=\linewidth}
\begin{tabular}{c|c|c|c}
\hline
\textbf{Tasks} & \textbf{Dataset} & \textbf{Language} & \textbf{\# of test instance} \\ \hline \hline
\multirow{6}{*}{CodSum} & \multirow{6}{*}{CSN \cite{husain2019codesearchnet}} & Python & 14,918 \\ \cline{3-4} 
 &  & PHP & 14,014 \\ \cline{3-4} 
 &  & Go & 8,122 \\ \cline{3-4} 
 &  & Java & 10,955 \\ \cline{3-4} 
 &  & JavaScript & 3,291 \\ \cline{3-4} 
 &  & Ruby & 1,261 \\ \hline \hline
\multirow{2}{*}{CodTran} & \multirow{2}{*}{CT \cite{lu2021codexglue}} & C\# & 1,000 \\ \cline{3-4} 
 &  & Java & 1,000 \\ \hline \hline
\multirow{2}{*}{CodGen} & HumanEval \cite{chen2021evaluating} & Python & 164 \\ \cline{2-4} 
 & MBPP \cite{austin2021program} & Python & 500 \\ \hline \hline
\end{tabular}
\end{adjustbox}
\end{table}

\subsection{Dataset and Data Process}
\label{sec:dataset}
We use well-known benchmarks for each examined task. The rationale for using these benchmarks is their commonality, enabling comparison with other studies without retraining models, and allowing assessment of generalization across multiple languages. For code summarization and translation, we use CodeXGLUE (CSN and CT)~\cite{lu2021codexglue}. 
For code generation, we use HumanEval\cite{chen2021evaluating} and MBPP~\cite{austin2021program}. Dataset statistics are in Table \ref{tab:dataset}. Since we do not train a new model, we show only the number of instances in the test sets used for evaluation.

Since \emph{GPT-4}'s training data is not public, it is uncertain if these benchmarks are included in its training. To reduce potential information leakage, we collected sample test sets from GitHub repositories created after Sept. 2021 (\emph{GPT-4}'s end of training) for qualitative analysis. The criteria for selecting repositories included: created after Oct. 2021, written in \texttt{Java}, actively maintained, not a fork, and with English comments. We selected three top-rated projects in different domains (library, algorithm, web) and picked two examples per task, totalling 18 sets of problems. The average token length is 14 for NL descriptions and 38 for SC snippets. The selected examples cover different domains, include multi-modalities and vary in difficulty. 

We followed common pre- and post-processing for each task. For code summarization, we removed special characters (except commas and periods) and tokenized them. For code generation, we formatted generated \texttt{Python} code to match the datasets. For code translation, we tokenized the generated code using \emph{ctok} in \texttt{Python}, as evaluation metrics are sensitive to tokenization. \emph{GPT-4} sometimes generated non-code text, requiring post-processing to keep only code. 
To automate \emph{GPT-4} prompting, we used the \emph{OpenAI} API\footnote{\url{https://platform.openai.com/}}, specifying the \emph{GPT-4} model. We set the temperature parameter to zero for better determinism \cite{ouyang2023llm}. Metrics were calculated using scripts from the benchmark dataset's repository. 

\subsection{Baselines}
\label{sec:baselines}
Table \ref{tab:baselines} lists the fine-tuned baseline models compared with \emph{GPT-4}. These are the best-performing fine-tuned models reported in each benchmark's leaderboard. Note that some baselines reported on the leaderboard were not publicly available, so we only included publicly available models in the table. The leaderboard can be found from each dataset: CodeXGLUE (CSN \& CT)\footnote{\url{https://microsoft.github.io/CodeXGLUE/}}, HumanEval\footnote{\url{https://paperswithcode.com/sota/code-generation-on-humaneval}}, and MBPP\footnote{\url{https://paperswithcode.com/sota/code-generation-on-mbpp}}.

\begin{table}[t!]
\centering
\caption{Baseline Fine-tuned models in three code tasks.}
\label{tab:baselines}
\begin{adjustbox}{width=\linewidth}
\begin{tabular}{|l|l|l|l|l|}
\hline
\textbf{Tasks} & \textbf{Baseline} & \textbf{Pre-trained Model} & \textbf{Param} & \textbf{Fine-Tuned} \\ \hline
\multirow{3}{*}{\begin{tabular}[c]{@{}l@{}}Code\\ Summarization\end{tabular}} & PolyglotCodeBERT \cite{ahmed2022multilingual} & CodeBERT \cite{feng2020codebert} & 125M & \multirow{3}{*}{\begin{tabular}[c]{@{}l@{}}CodeSearchNet\\ (CSN)\end{tabular}} \\ \cline{2-4}
 & CoTexT \cite{phan2021cotext} & T5-base \cite{raffel2020exploring} & 220M &  \\ \cline{2-4}
 & ProphetNet-X \cite{qi2021prophetnet} & ProphetNet-Code \cite{qi2021prophetnet} & 300M &  \\ \hline
\multirow{7}{*}{\begin{tabular}[c]{@{}l@{}}Code \\ generation\end{tabular}} & PanGu-Coder2 \cite{shen2023pangu} & StarCoder \cite{listarcoder} & 15B & \multirow{3}{*}{\begin{tabular}[c]{@{}l@{}}HumanEval\\ (HE)\end{tabular}} \\ \cline{2-4}
 & WizardCoder \cite{luowizardcoder} & StarCoder \cite{listarcoder} & 15B &  \\ \cline{2-4}
 & XCoder \cite{wang2024your} & LLaMA3-8B-Base \cite{dubey2024llama} & 8B &  \\ \cline{2-5} 
 & CODE-T-ITER \cite{chen2022codet} & \multirow{2}{*}{\begin{tabular}[c]{@{}l@{}}cushman-001\\ davinci-001\&002 \cite{chen2021evaluating} \end{tabular}} & \multirow{2}{*}{\begin{tabular}[c]{@{}l@{}}12B\\ 175B\end{tabular}} & HE \& MBPP \\ \cline{2-2} \cline{5-5} 
 & CODE-T \cite{chen2022codet} &  &  & \multirow{3}{*}{MBPP} \\ \cline{2-4}
 & StarCoder \cite{listarcoder} & StarCoderBase \cite{listarcoder} & 15.5B &  \\ \cline{2-4}
 & StarCoder2 \cite{lozhkov2024starcoder} & StarCoder2-15B \cite{lozhkov2024starcoder} & 15B &  \\ \hline
\multirow{2}{*}{\begin{tabular}[c]{@{}l@{}}Code\\ translation\end{tabular}} & StructCoder \cite{tipirneni2023structcoder} & CodeT5 based \cite{wang2021codet5} & 224M & \multirow{2}{*}{\begin{tabular}[c]{@{}l@{}}CodeTrans\\ (CT)\end{tabular}} \\ \cline{2-4}
 & PLBART \cite{ahmad2021unified} & PLBART \cite{ahmad2021unified} & 140M &  \\ \hline
\end{tabular}
\end{adjustbox}
\end{table}

\subsection{Evaluation Metrics}
\label{sec:metrics}
We use the same metrics as the benchmarks to evaluate \emph{GPT-4}. For code generation, we use $pass@k$~\cite{chen2021evaluating}, defined as the probability that one of the top k-generated samples passes the unit tests (with k set to 1). For code summarization, we use $BLEU$ \cite{papineni2002bleu}, which assesses similarity to the ground truth using n-gram precision. For code translation, we use $BLEU$, $ACC$~\cite{stehman1997selecting}, and $CodeBLEU$~\cite{ren2020codebleu}, which combines n-gram precision, keyword matching, AST matching, and dataflow matching.

\begin{figure}[t!]
    \centering
     {\includegraphics[width=\linewidth]{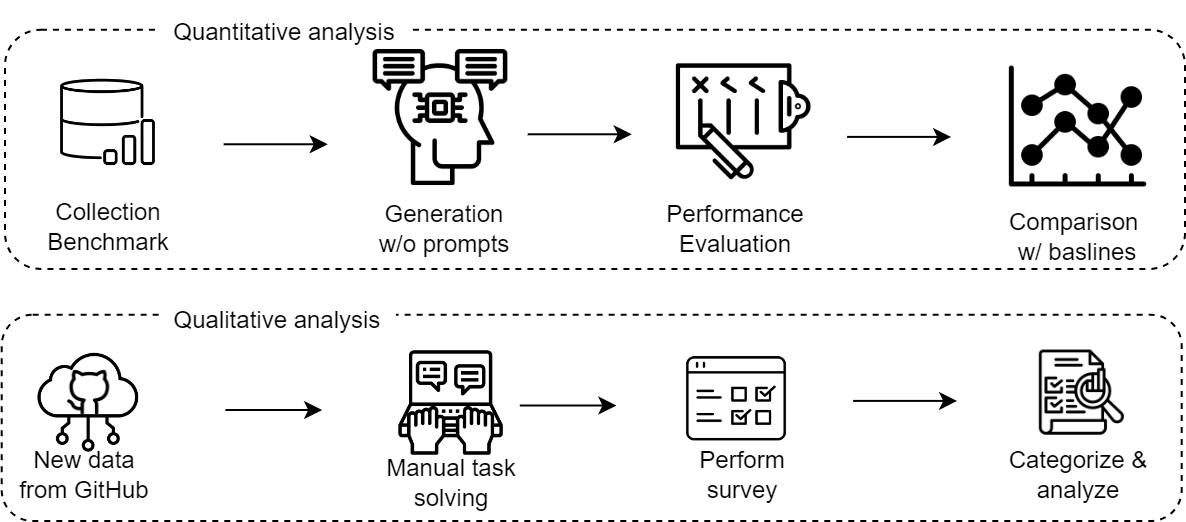}}
     \caption{Overall workflow of our study.}
    \label{fig:overview}
\end{figure}

\section{Methodology and Protocols}
\label{sec:emp_method}
In this section, we discuss the details of the methodology and the protocol designs for our quantitative and qualitative study.
Figure \ref{fig:overview} shows the overall workflow of this study.


\subsection{Quantitative Study of \emph{GPT-4} with Baselines (RQ1)}
\label{sec:quatitative}

To quantitatively assess the performance of \emph{GPT-4} for each ASE task, we use widely known and used benchmark datasets: CodeXGLUE, HumanEval, and MBPP (details in Section \ref{sec:dataset}). 
For the baseline on fine-tuned models, we report the scores from the leaderboard of each benchmark and compare them with the evaluation metrics calculated from the results of \emph{GPT-4}.
To generate the results from \emph{GPT-4}, we use three different kinds of prompting strategies:

\begin{enumerate} 
    \item \textbf{Basic prompting \cite{brown2020language,liu2023pre}:} we directly query \emph{GPT-4} with the input (code or description) and ask to generate solutions in the form of the desired output.
    \item \textbf{In-context prompting \cite{liu2022makes,geng2024large}:} together with the basic prompt, we give a set of input/output examples to \emph{GPT-4}. This idea is very similar to few-shot learning in the context of fine-tuned language models.
    \item  \textbf{Task specific engineered prompting \cite{he2022hyperprompt,carta2023iterative,liu2023improving}:} together with the basic prompt, we design additional prompts to guide \emph{GPT-4} in generating better results for each task.
\end{enumerate}
{We have selected three prompting strategies covering a range of prompts from simple to more advanced. We use a small subset (between 10 and 50, depending on the task) to evaluate how a crafted prompt results from the actual data (using the BLEU metric). This phase is done to avoid obvious mistakes and not to optimize the prompt for each task. Note that finding the optimal prompting strategy is not the goal of the paper. Although the selections of prompts do not cover all possibilities, our goal is to provide an initial study on automated prompting techniques in three different levels of complexity. We select basic prompting as a baseline or the least complex, in-context learning as one of the best and most generic prompting approaches (at the time of conducting this study) and task-specific prompting as a specialized ad-hoc prompt that potentially outperforms in-context learning. We make the basic prompt as simple and generic as possible to compare how adding different strategies shows/improves different results. Similarly, for in-context learning, we want to keep the prompt as close to the basic prompt as possible and only add different contexts so we can compare how giving examples improves the basic prompts.} 

\subsubsection{Basic Prompting}
\label{sec:basic_prompting}
The basic prompts used for each task are shown in Listing \ref{lst:basic_prompt}.
They are the simplest prompts that consist of the input code/description and the description of the expected output.

\definecolor{dkgreen}{rgb}{0,0.6,0}
\definecolor{gray}{rgb}{0.5,0.5,0.5}
\definecolor{mauve}{rgb}{0.58,0,0.82}
\lstset{frame=tb,
  language=Python,
  aboveskip=3mm,
  belowskip=3mm,
  showstringspaces=false,
  columns=flexible,
  basicstyle={\scriptsize\ttfamily},
  numbers=none,
  numberstyle=\scriptsize\color{gray},
  keywordstyle=\color{blue},
  commentstyle=\color{dkgreen},
  stringstyle=\color{mauve},
  breaklines=true,
  breakatwhitespace=true,
  tabsize=3
}
\vspace{10pt}
\begin{minipage}{0.9\linewidth}
\begin{lstlisting}[caption={Basic prompt used to query.},captionpos=b,label={lst:basic_prompt}]
### code summarization task
f"Given a {lang} code snippet surrounded in ???, generate one line of semantic focused and abstract summarization ???{code}???"

### code generation task
f"Generate Python source code for a function, given
a natural language prompt (surrounded by ???)
describing the function's purpose. Output only the
Python source code on exactly one line. ???{nl}???"

### code translation task
f"Given that the code surrounded by ??? is written in C#, output only the corresponding Java code condensed on one line ???{code}???"

\end{lstlisting}
\end{minipage}

\subsubsection{In-Context Prompting}
\label{sec:icl_prompting}
For in-context prompting, we add input/output examples on the basic prompt as additional contexts.
In-context prompting is one of the advanced types of automated prompts that help LLM learn the additional context.
We select three input-output examples for the context.
A very high number of examples will consume more tokens that might not fit into the allowed context length and also be more expensive. 
To select the context examples, we follow the existing studies \cite{gao2023constructing,yuan2023evaluating} that used BM25 similarity-based selection \cite{robertson2009probabilistic}.
We calculate the similarity of each test instance's input to each training instance's input and select the top three similar instances for the context examples.
The training set here refers to the benchmark datasets, i.e., the dataset in Table. \ref{tab:dataset}.
Although we do not use them for training, we use them to select the context examples.
It was shown to outperform the random selection or the fixed selection for in-context learning \cite{gao2023constructing,yuan2023evaluating}.  

\subsubsection{Task-Specific Engineered Prompts}
\label{sec:task_specific}
We created task-specific prompts using simple heuristics to address the challenges we faced with basic and in-context prompting strategies. We manually analyzed a subset of outputs to identify the root causes of issues. We discuss the challenges and how we addressed them with task-specific prompts in the following paragraphs.

For code summarization, we tested \emph{GPT-4}'s initial responses using ten samples and found that the BLEU score decreased due to verbose explanations. To improve the score, we instructed \emph{GPT-4} to limit the output to 15 tokens, the average token length of the training set's outputs.
We also found cases where \emph{GPT-4} tries to summarize code for each line of code at the syntax level.
To mitigate this, we added ``generate one line of semantic focused and abstract summary of the code'' to the prompt.
Also, due to the creativity and immense dictionary of \emph{GPT-4}, it generates synonyms that are not used in the input.
To mitigate this problem, we added instructions to guide the model in using naturalized identifiers to form the summary.

For code generation, we found that \emph{GPT-4} had a hard time inferring the steps that were missing in the input (description of code). To mitigate this, we tried to guide \emph{GPT-4} to consider the steps required to solve the described problem carefully. First, we selected five potential prompts. Two of them used in-context prompting, one used chain-of-thought \cite{wei2022chain}, and the final two prompts combined in-context and chain-of-thought. For the prompts with in-context prompting, we randomly selected three samples from the sanitized train set for the MBPP dataset and the test set for the HumanEval dataset. For the samples, we selected 50 examples out of 120 from the sanitized train set for the MBPP dataset and 50 randomly selected examples out of 164 from the test set for the HumanEval dataset. Based on the Pass@1 scores, we chose the best-performing prompt.
From the preliminary evaluation, we found that using only the chain-of-thought resulted in the best performance.

For code translation, we selected ten samples to check the initial results.
{We observed that \emph{GPT-4} fails to generate syntactic keywords that were prevalent in the ground truth, e.g., \texttt{virtual} and \texttt{override}. However, missing syntactic keywords were easily mitigated by giving more context to the different languages, i.e., input-output examples.} Therefore, we decided to modify the in-context prompt to overcome its weaknesses. Five code pair examples were chosen as opposed to three for the in-context prompt, with the rationale being that a wider variety of lengths and keywords would provide better exposure to the ``quirks'' of the dataset that the first in-context prompt missed (e.g., \texttt{C\#} code referring to \texttt{Java} libraries such as \texttt{java.nio.ByteBuffer}).
Listing \ref{lst:task_specific_prompt} shows the resulting task-specific prompt.

\definecolor{dkgreen}{rgb}{0,0.6,0}
\definecolor{gray}{rgb}{0.5,0.5,0.5}
\definecolor{mauve}{rgb}{0.58,0,0.82}

\lstset{frame=tb,
  language=Python,
  aboveskip=3mm,
  belowskip=3mm,
  showstringspaces=false,
  columns=flexible,
  basicstyle={\scriptsize\ttfamily},
  numbers=none,
  numberstyle=\tiny\color{gray},
  keywordstyle=\color{blue},
  commentstyle=\color{dkgreen},
  stringstyle=\color{mauve},
  breaklines=true,
  breakatwhitespace=true,
  tabsize=3
}
\vspace{10pt}
\begin{minipage}{0.9\linewidth}
\begin{lstlisting}[caption={Task-specific prompt used to query.},captionpos=b,label={lst:task_specific_prompt}]
### code summarization task
f"Generate one line of semantic focused and abstract summary of the code surrounded by ???. Compose the summarization by naturalizing the identifier of variables and function names in the code as keywords. The summarization should be very concise, with an approximate limitation of around 15 tokens in length."

### code generation task
f"Generate Python source code for a function, given a natural language prompt (surrounded by ???) describing the function's purpose. Carefully consider the steps required to fulfill the function's purpose stated in the natural language prompt. Output only the Python source code on exactly one line. ???{nl}???"

### code translation task
f"Given the five examples of translating Java code to C# code (both surrounded by ???), translate the 5 Java code samples (surrounded by ???) to C# code and output each C# code on exactly one line. Do not include any extra text such as 'C# code sample', and do not include surrounding ??? either. ???{5_i_o_examples}???"
\end{lstlisting}
\end{minipage}

\subsection{Qualitative Analysis of Participants' Perception and Interaction with GPT-4 (RQ2 and RQ3)}
\label{sec:qualitative}
To answer RQ2 and RQ3, we performed a user study by surveying participants from both academia and industry.
We first discuss the protocols for sampling and conducting this empirical evaluation.

\begin{figure}[t!]
    \centering
    \begin{subfigure}[b]{\linewidth}
        \includegraphics[width=\linewidth]{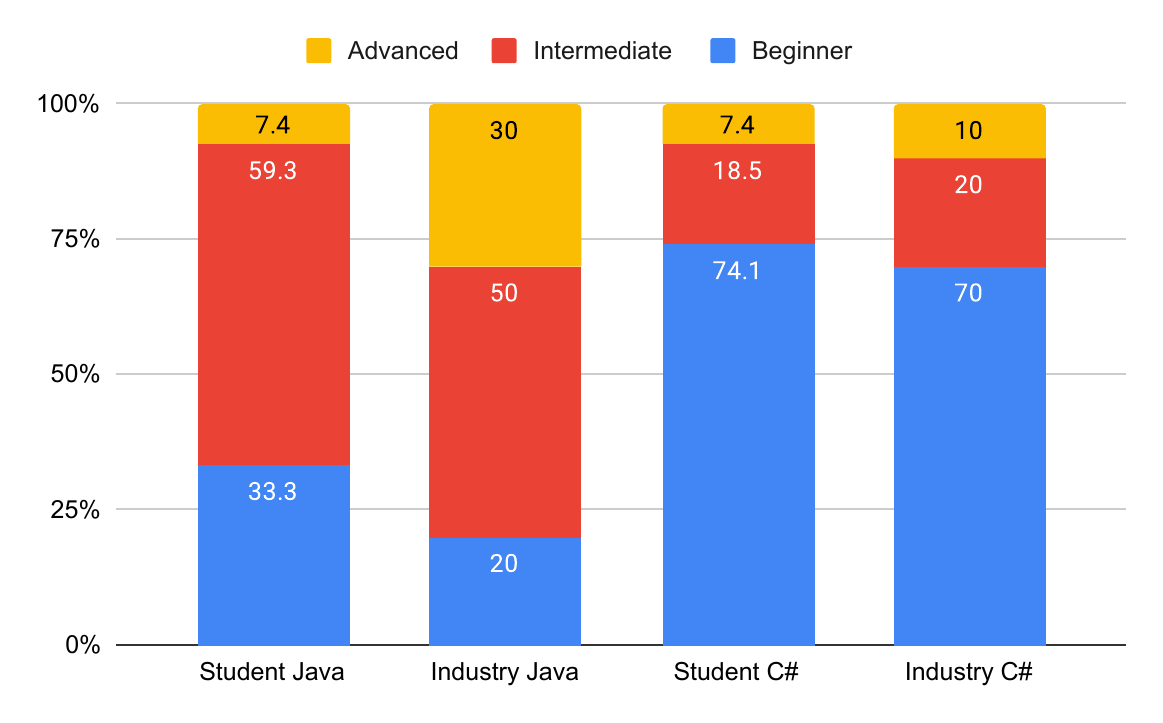}
        \subcaption{Participant programming language proficiency.}
        \label{fig:demographics_a}
    \end{subfigure}
    \begin{subfigure}[b]{\linewidth}
        \includegraphics[width=\linewidth]{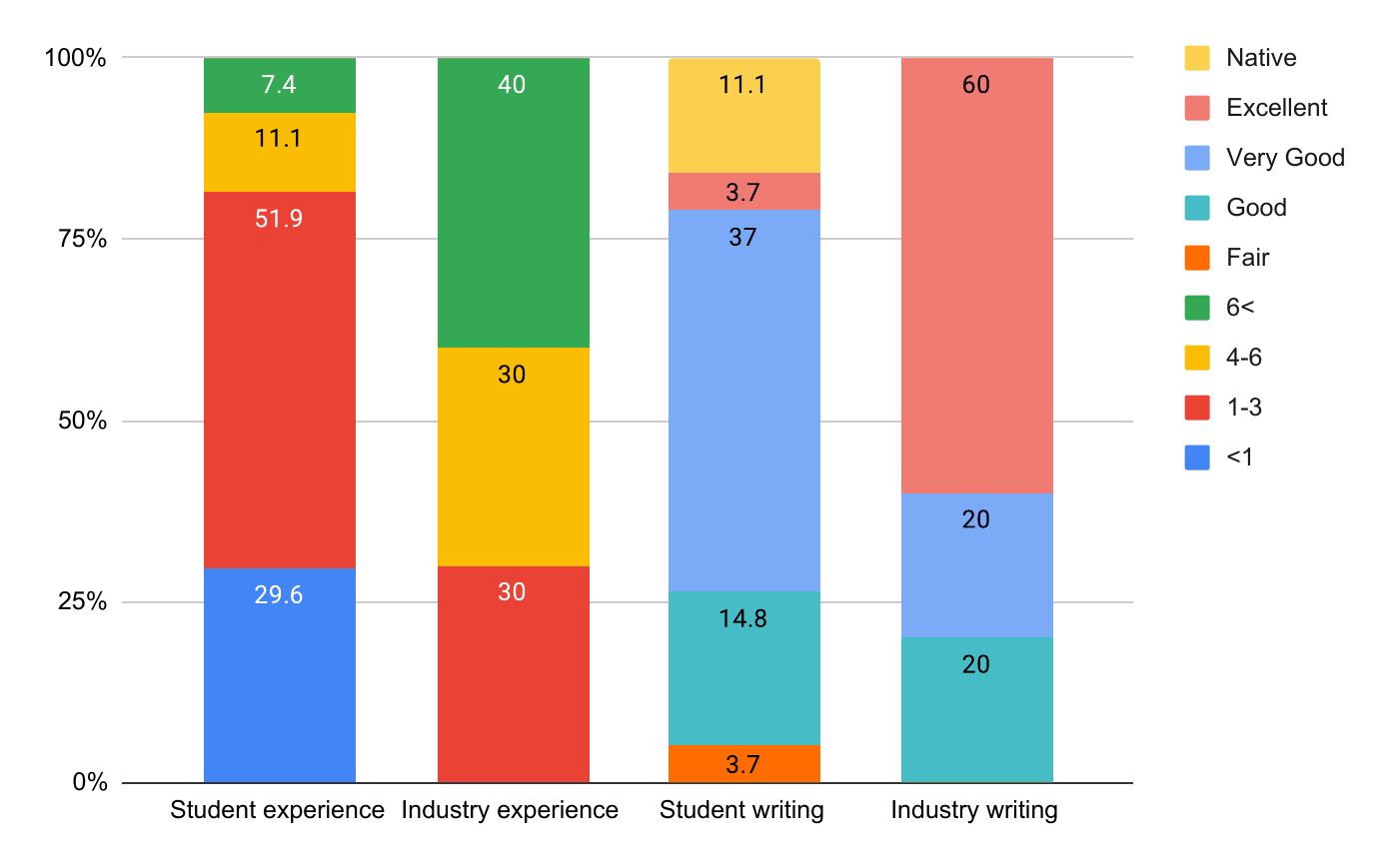}
        \subcaption{Programming experience and writing proficiency.}
        \label{fig:demographics_b}
    \end{subfigure}
    \caption{Demographics from the survey.}
    \label{fig:demographics}
\end{figure}
\noindent\textit{\textbf{Survey Participants:}}
{
We conducted a user study with 27 graduate students (recruited via convenience sampling \cite{kitchenham2002principles} from an EECS mailing list; 31 respondents, 27 selected for availability) and 10 industry professionals (recruited from 26 invited across Canadian, U.S., and Asian tech companies; 12 respondents, 10 selected).}

{
Participants’ demographics are shown in Figure \ref{fig:demographics}.
From Academic participants, the distributions are 77.8\% M.Sc., 18.5\% Ph.D., 1 Postdoc.
From Industry participants, the distributions are 4 senior software engineers, 3 software engineers, 1 associate ML developer, 1 lead engineer, and 1 MLOps lead.
For LLM familiarity, 5.4\% never used, 37.8\% application-level users, 29.7\% users with blog/research exposure, 8.1\% implemented models, 16.2\% active researchers, 2.7\% no prior exposure.}

\noindent\textit{\textbf{Study Design:}} The survey consists of three primary phases, i.e., the {background and demographic survey}, {user observation}, and {post-experimentation survey}.

\noindent \textit{\textbf{Background and Demographic:}}
We ask for participants' demographic information, i.e., years of experience in programming outside of school, degree of familiarity with LLMs, current position, degree of familiarity with the target language (\texttt{Java} and \texttt{C\#}), and the level of technical writing skills in English. We asked six close-ended questions overall. 

\noindent \textit{\textbf{User Observation Study:}}
We assign two out of three tasks to each participant, considering their familiarity with programming languages.
Since participants should have some knowledge about \texttt{C\#} for the code translation task, we first asked if they knew \texttt{C\#} and only gave them the code translation task.
The rest of the tasks were assigned randomly.
For each task, we gave one sample from each of 3 projects.
The participants worked on the task in a conversation with \emph{ChatGPT-4}.
\emph{ChatGPT-4} is a chat-interfaced framework that runs on \emph{GPT-4} \footnote{\url{https://chat.openai.com/}}.
We keep the settings of \emph{ChatGPT-4} to default.
We ensured a balanced allocation of task samples to participants randomly (with an equal number of participants per project and task).
We ask participants to start with their basic prompt and then to proceed with conversations to improve the result based on their first evaluation until they consider the solution satisfactory or have spent 5 minutes per task sample.
In this setup, we have not provided any example prompts to avoid any bias and to get participants' fresh perspectives.
Overall, we evaluated three tasks over six samples per task.
Each sample has been assigned and solved with an average of 12 times by different participants \cite{guest2006many}. 
The survey was designed to take no more than 60 minutes for each participant.

\noindent \textit{\textbf{Post-Experimentation Survey:}} After the observation study, we posed four closed-ended questions to the participants.
We inquire whether they received an incorrect response from the model, the degree to which they found \emph{GPT-4} useful for the assigned task, their perception of response quality on the first and last attempt, and their overall assessment. Additionally, we included four open-ended questions to assess the utility of the \emph{GPT-4} model.
Strengths and weaknesses of the first and last response for both tasks.

For RQ2, we report the responses from the four close-ended questions explained in the post-experimentation survey.
We show the percentage of the evaluation responses from all the participants.
The evaluation of each survey question can be answered to one out of the five following selections: 1) poor, 2) fair, 3) good, 4) very good, and 5) excellent.
We consider the evaluation to be satisfactory if they answer either good, very good, or excellent.

For RQ3, we investigate the chat logs of the participants and manually categorize the conversational prompts.
{We first extracted the conversational prompts as a list of abstract instructions for the qualitative analysis. We then used an Open Coding Approach, where two authors of this paper, who were not involved in the extraction process, performed sessions of card sorting for our qualitative data analysis~\cite{miles1994qualitative}. This analysis design was chosen to make the categorization and extraction as independent from each other as possible. After that, a third author acted as a moderator to resolve the disagreements. Finally, all other authors are involved in manually reviewing the forms completed by the three authors and resolving any disagreements that occurred during the process to ensure consensus among reviewers. Note that we also provide the category of conversational prompt and the raw chat logs for each participant in our replication package to provide transparent results of our work.}
After categorizing the types of conversational prompts, we count and rank the occurrences for each task to show the overall trends among the participants.

\subsection{Analysis of Users' Conversational Prompts (RQ4)}
For RQ4, we investigate each prompt created by the participants and the response by \emph{GPT-4}. 
Since \emph{GPT-4} generates verbose explanations for its response, we manually extract the core part of the answer for each task.
For \textit{code generation} and \textit{translation tasks}, we excerpt the code snippets from the generated response. 
For \textit{code summarization tasks}, we excerpt the texts that summarize the given code snippet.
The responses are written in a specific code editor format, allowing us to extract them from the logs manually.
After we extract the answers, we rank them based on the BLEU score. 
For each category of conversational prompt, we calculate the mean reciprocal rank (MRR) to find which type of prompt constitutes the most to a higher rank. 
Lastly, we report the average BLEU score of the conversational prompts, the highest BLEU score achieved by a participant, and the generation from the two automated prompt strategies we used in our quantitative study, i.e., basic prompts and tasks-specific prompts, to compare the performance of conversational prompts. 
We did not include the in-context prompt as a comparison baseline for this experiment as we do not have a separate training set to select similar context examples (as mentioned in Section \ref{sec:icl_prompting}). 

All the survey questions, survey responses, the raw chat logs from the participants, and analyzed prompt categories by the authors are available in our replication package.

\section{Experimental Results}
\label{sec:results}

\subsection{RQ1: Automated Prompts vs. Fine-Tuned Models}
\label{sec:rq1_result}

\begin{table}[t!]
\centering
\caption{Result of code summarization. Bold denotes the highest score. The asterisk (*) denotes that the source was not public. GPT-4 + B/IC/TS denotes basic, in-context, and task-specific prompts, respectively.}
\label{tab:rq1_comment}
\begin{adjustbox}{width=\linewidth}
\begin{tabular}{l|ccccccc|}
\cline{2-8}
 & \multicolumn{7}{c|}{\textbf{BLEU}} \\ \hline
\multicolumn{1}{|l|}{\textbf{Model}} & \multicolumn{1}{l|}{\textbf{Ruby}} & \multicolumn{1}{c|}{\textbf{JS}} & \multicolumn{1}{c|}{\textbf{Go}} & \multicolumn{1}{c|}{\textbf{Py}} & \multicolumn{1}{l|}{\textbf{Java}} & \multicolumn{1}{l|}{\textbf{PHP}} & \textbf{Avg} \\ \hline
\multicolumn{1}{|l|}{StarCoder-LoRA*} & \multicolumn{1}{l|}{17.21} & \multicolumn{1}{l|}{18.15} & \multicolumn{1}{c|}{21.61} & \multicolumn{1}{c|}{23.13} & \multicolumn{1}{l|}{22.61} & \multicolumn{1}{l|}{28.76} & 21.91 \\ \hline 
\multicolumn{1}{|l|}{DistillCodeT5*} & \multicolumn{1}{l|}{15.75} & \multicolumn{1}{l|}{16.42} & \multicolumn{1}{c|}{20.21} & \multicolumn{1}{c|}{20.59} & \multicolumn{1}{l|}{20.51} & \multicolumn{1}{l|}{26.58} & 20.01 \\ \hline
\multicolumn{1}{|l|}{PolyglotCodeBERT \cite{ahmed2022multilingual}} & \multicolumn{1}{l|}{14.75} & \multicolumn{1}{l|}{15.80} & \multicolumn{1}{c|}{18.77} & \multicolumn{1}{c|}{18.71} & \multicolumn{1}{c|}{20.11} & \multicolumn{1}{l|}{26.23} & 19.06 \\ \hline
\multicolumn{1}{|l|}{CoTexT \cite{phan2021cotext}} & \multicolumn{1}{l|}{14.02} & \multicolumn{1}{l|}{14.96} & \multicolumn{1}{c|}{18.86} & \multicolumn{1}{c|}{19.73} & \multicolumn{1}{l|}{19.06} & \multicolumn{1}{l|}{24.68} & 18.55 \\ \hline
\multicolumn{1}{|l|}{ProphetNet-X \cite{qi2021prophetnet}} & \multicolumn{1}{l|}{14.37} & \multicolumn{1}{l|}{16.60} & \multicolumn{1}{c|}{18.43} & \multicolumn{1}{c|}{17.87} & \multicolumn{1}{l|}{19.39} & \multicolumn{1}{l|}{24.57} & 18.54 \\ \hline \hline
\multicolumn{1}{|l|}{{GPT-4 +} B} & \multicolumn{1}{l|}{19.93} & \multicolumn{1}{l|}{19.96} & \multicolumn{1}{l|}{29.70} & \multicolumn{1}{c|}{27.19} & \multicolumn{1}{l|}{24.11} & \multicolumn{1}{l|}{18.63} & 23.25 \\ \hline
\multicolumn{1}{|l|}{{GPT-4 +} IC } & \multicolumn{1}{l|}{19.60} & \multicolumn{1}{l|}{20.08} & \multicolumn{1}{l|}{28.98} & \multicolumn{1}{c|}{19.90} & \multicolumn{1}{l|}{24.45} & \multicolumn{1}{l|}{19.24} & 22.04 \\ \hline
\multicolumn{1}{|l|}{{GPT-4 +} TS} & \multicolumn{1}{l|}{\textbf{25.48}} & \multicolumn{1}{l|}{\textbf{27.23}} & \multicolumn{1}{l|}{\textbf{35.34}} & \multicolumn{1}{c|}{\textbf{29.61}} & \multicolumn{1}{l|}{\textbf{33.08}} & \multicolumn{1}{l|}{\textbf{30.67}} & \textbf{30.24} \\ \hline
\end{tabular}
\end{adjustbox}
\end{table}

Table \ref{tab:rq1_comment} shows the results of fine-tuned baselines and \emph{GPT-4} on the code summarization task. 
The basic prompting improved the 1st ranked baseline by 1.34\% points on average.
However, this pattern wasn't consistent across all programming languages where for \texttt{PHP}, we see a 10.13\% points decrease. We observe the biggest improvement from \texttt{Go} with 8.09\% points improvement. 
The in-context prompting improved the 1st ranked baseline by 0.13\% points on average. However, the inconsistent pattern was similar to the basic prompt where a decrease was shown in \texttt{PHP} with a 9.52\% points and the highest improvement from \texttt{Go} with a 7.37\% points improvement. 
When comparing in-context prompting to basic prompting, it didn't improve the basic prompting, showing a 1.21\% points decrease on average.
The lowest decrease was found in \texttt{Python} with 7.29\% points. The highest improvement was found in \texttt{PHP} with 0.61\% points. 
The task-specific prompt had the most noticeable improvement compared to the 1st ranked baseline with 8.33\% points improvement on average.
When comparing task-specific to the basic prompt, it improved by 6.99\% points overall.

\begin{table}[t!]
\centering
\caption{Result of code generation task.} 
\label{tab:rq1_code}
\begin{adjustbox}{width=\linewidth}
\begin{tabular}{|cc|cc|}
\hline
\multicolumn{2}{|c|}{\textbf{HumanEval}} & \multicolumn{2}{c|}{\textbf{MBPP}} \\ \hline
\multicolumn{1}{|l|}{\textbf{Model}} & \textbf{Pass@1} & \multicolumn{1}{l|}{\textbf{Model}} & \textbf{Pass@1} \\ \hline
\multicolumn{1}{|l|}{CODE-T (davinci-002) \cite{chen2022codet}} & 65.80 & \multicolumn{1}{l|}{CODE-T (davinci-002)\cite{chen2022codet}} & \textbf{67.70} \\ \hline
\multicolumn{1}{|l|}{CODE-T-Iter (davinci-002) \cite{chen2022codet}} & 65.20 & \multicolumn{1}{l|}{StarCoder2 \cite{lozhkov2024starcoder}} & 66.20 \\ \hline
\multicolumn{1}{|l|} {PanGu-Coder2 \cite{shen2023pangu}} & 61.64 & \multicolumn{1}{l|}{CODE-T (davinci-001)\cite{chen2022codet}} & 61.90\\ \hline
\multicolumn{1}{|l|}{WizardCoder\cite{luowizardcoder}}  & 57.30	 & \multicolumn{1}{l|}{CODE-T (cushman-001)\cite{chen2022codet}} & 55.40 \\ \hline
\multicolumn{1}{|l|}{XCoder \cite{wang2024your}} & 57.30 & \multicolumn{1}{l|}{StarCoder \cite{listarcoder}} & 52.70 \\ \hline \hline
\multicolumn{1}{|l|}{{GPT-4 +} B} & 69.51 & \multicolumn{1}{l|}{{GPT-4 +} B} & 39.60 \\ \hline
\multicolumn{1}{|l|}{{GPT-4 +} IC} & 64.63 & \multicolumn{1}{l|}{{GPT-4 +} IC } & 40.60 \\ \hline
\multicolumn{1}{|l|}{{GPT-4 +} TS} & \textbf{74.39} & \multicolumn{1}{l|}{{GPT-4 +} TS} & 39.40 \\ \hline
\end{tabular}
\end{adjustbox}
\end{table}

Table \ref{tab:rq1_code} shows the results for the code generation tasks.
The basic prompting was able to improve the 1st ranked baseline with an increase of 3.71\% points for HumanEval.
However, for MBPP, it decreased by 28.10\% points. 
The in-context prompting also couldn't outperform the baselines (1.17\% points decrease for HumanEval and 27.64\% points decrease for MBPP). 
When comparing in-context prompting to basic prompting, there is a decrease of 4.88\% points in HumanEval and a small increase of 1.00\% points in MBPP. 
The task-specific prompting improved from the baseline, with an 8.59\% points increase in HumanEval.
However, for MBPP, it decreased by 28.30\% points.
When comparing task-specific to basic prompting, it didn't show a significant improvement, showing a 4.88\% points improvement in HumanEval and a decrease of 0.20\% points in MBPP.


\begin{table}[t!]
\centering
\caption{Result of code translation. Bold denotes the highest score. The asterisk (*) denotes that the source was not public. GPT-4 + B/IC/TS denotes basic, in-context, and task-specific prompts, respectively.}
\label{tab:rq1_translation}
\begin{adjustbox}{width=\linewidth}
\begin{tabular}{l|ccc|ccc|}
\cline{2-7}
 & \multicolumn{3}{c|}{\textbf{Java to C\#}} & \multicolumn{3}{c|}{\textbf{C\# to Java}} \\ \hline
\multicolumn{1}{|l|}{\textbf{Model}} & \multicolumn{1}{l|}{\textbf{BLEU}} & \multicolumn{1}{l|}{\textbf{ACC}} & \multicolumn{1}{l|}{\textbf{CB}} & \multicolumn{1}{l|}{\textbf{BLEU}} & \multicolumn{1}{l|}{\textbf{ACC}} & \multicolumn{1}{l|}{\textbf{CB}} \\ \hline
\multicolumn{1}{|l|}{StructCoder \cite{tipirneni2023structcoder}} & \multicolumn{1}{r|}{\textbf{85.02}} & \multicolumn{1}{r|}{\textbf{66.60}} & \textbf{88.42} & \multicolumn{1}{r|}{80.66} & \multicolumn{1}{r|}{\textbf{67.70}} & 86.03 \\ \hline
\multicolumn{1}{|l|}{PLNMT-sys0*} & \multicolumn{1}{r|}{83.37} & \multicolumn{1}{r|}{64.60} & 87.38 & \multicolumn{1}{r|}{80.91} & \multicolumn{1}{r|}{66.80} & 85.87 \\ \hline
\multicolumn{1}{|l|}{PLBART \cite{ahmad2021unified}} & \multicolumn{1}{r|}{83.02} & \multicolumn{1}{r|}{64.60} & 87.92 & \multicolumn{1}{r|}{78.35} & \multicolumn{1}{r|}{65.00} & 85.27 \\ \hline
\multicolumn{1}{|l|}{CodePALM*} & \multicolumn{1}{r|}{83.26} & \multicolumn{1}{r|}{65.50} & 86.37 & \multicolumn{1}{r|}{78.94} & \multicolumn{1}{r|}{65.20} & 83.74 \\ \hline
\multicolumn{1}{|l|}{ContraBERT\_G*} & \multicolumn{1}{r|}{80.78} & \multicolumn{1}{r|}{59.90} & 84.98 & \multicolumn{1}{r|}{76.24} & \multicolumn{1}{r|}{60.50} & 81.69 \\ \hline \hline
\multicolumn{1}{|l|}{{GPT-4 +} B} & \multicolumn{1}{r|}{55.33} & \multicolumn{1}{r|}{5.50} & 68.34 & \multicolumn{1}{r|}{63.08} & \multicolumn{1}{r|}{20.20} & 73.57 \\ \hline
\multicolumn{1}{|l|}{{GPT-4 +} IC} & \multicolumn{1}{r|}{80.72} & \multicolumn{1}{r|}{34.10} & 86.69 & \multicolumn{1}{r|}{\textbf{84.12}} & \multicolumn{1}{r|}{50.20} & \textbf{87.44} \\ \hline
\multicolumn{1}{|l|}{{GPT-4 +} TS} & \multicolumn{1}{r|}{80.55} & \multicolumn{1}{r|}{34.30} & 86.87 & \multicolumn{1}{r|}{81.29} & \multicolumn{1}{r|}{49.20} & 85.33 \\ \hline
\end{tabular}
\end{adjustbox}
\end{table}

Table \ref{tab:rq1_translation} shows the results for code translation. 
The basic prompting couldn't outperform the 1st ranked baseline with a decrease of 29.69\% points in BLEU, 61.10\% points in ACC, and 20.08\% points in CodeBLEU for translating \texttt{Java-C\#}.
Similar patterns showed when translating \texttt{C\#-Java} with a decrease of 17.83\% points in BLEU, 47.50\% points in ACC, and 12.46\% points in CodeBLEU.
The low ACC scores when translating \texttt{Java-C\#} are mainly caused by the failure to predict the frequently used syntactic keywords (e.g., \texttt{virtual} and \texttt{override}). 
For example, the keyword \texttt{override} appeared in 216, and \texttt{virtual} appeared in 431 out of the 1,000 test samples (totaling 647, as no sample used both keywords). 
Conversely, \emph{GPT-4} generated the keyword \texttt{override} in only 48 results and never generated \texttt{virtual}.  
We also observed that it failed to infer code elements that are missing from the input, e.g., parameters in API calls that are not in the source language but are in the target language.
Since ACC requires exact matching, a single token could prevent an exact match, severely impacting the overall ACC score. 
We observed another type of failure which was caused by the LLM hallucination \cite{alshahwan2024automated,yang2023harnessing,chen2023hallucination,guerreiro2023hallucinations} where the model generates content that is irrelevant, made-up, or inconsistent from the input data. For example, the model would generate APIs that are not used in the source language or even APIs that don't exist.
For \texttt{C\#-Java}, we did not find any noticeable frequently omitted syntactic keyword, which could explain the relatively higher ACC score.
A drastic improvement was achieved by in-context and task-specific prompting, especially for generating the missing syntactic keywords.
Overall, \emph{GPT-4} with in-context prompting showed mixed performance, i.e., it couldn't outperform the baselines in translating \texttt{Java-C\#} while there was an improvement in \texttt{C\#-Java}. 
When comparing in-context prompting and basic prompting, it showed a drastic improvement in both \texttt{Java-C\#} and \texttt{C\#-Java}. 
We observed that task-specific prompting outperforms basic prompting.
However, it could not outperform in-context learning, as we found in our initial subset when designing the prompt.

Overall, \emph{GPT-4} with different automated prompting strategies could not outperform the fine-tuned models significantly. For code summarization, a reason for this can be due to the verbose and creative generation of \emph{GPT-4}. Since the evaluation metrics heavily rely on textual similarity, diverse and verbose generation will significantly impact the metric scores. When we tried to mitigate this with our task-specific prompt, the result improved by a big margin, outperforming the fine-tuned baselines. For code generation, the code generated by the \emph{GPT-4} seems plausible and executable. However, they fail to pass the tests, e.g., off-by-one error, using imprecise logical operators (choosing between $and$ and $or$ operator), or missing to generate intermediate steps that are not mentioned in the NL description. A possible reason is that since it is not fine-tuned, it is harder to generate code that requires project-/data-specific knowledge, which could be leveraged to generate code with correct functionality. For code translation, the model failed to generate the same code (low ACC) for both target languages. The possible reasons can be: 1) a much larger and diverse corpus is used for training, resulting in a more creative generation, and 2) generating something irrelevant to the input, i.e., LLM hallucination. Failing to generate similar (low BLEU) \texttt{C\#} code compared to \texttt{Java} code can be due to the difference in data size used for training \emph{GPT-4}, since \texttt{Java} is more commonly used than \texttt{C\#} in GitHub\footnote{\url{https://gitnux.org/github-languages-statistics/}}.

\mybox{\textbf{Answer to RQ1}: Automated prompting with \emph{GPT-4} did not significantly outperform fine-tuned baselines, except for code summarization, indicating the need for more sophisticated prompting strategies to fully leverage \emph{GPT-4}'s capabilities.}

\subsection{RQ2: Participants' Perception of \emph{GPT-4}}
\label{sec:rq2_result}
\begin{figure}[t!]
    \centering
     {\includegraphics[width=\linewidth]{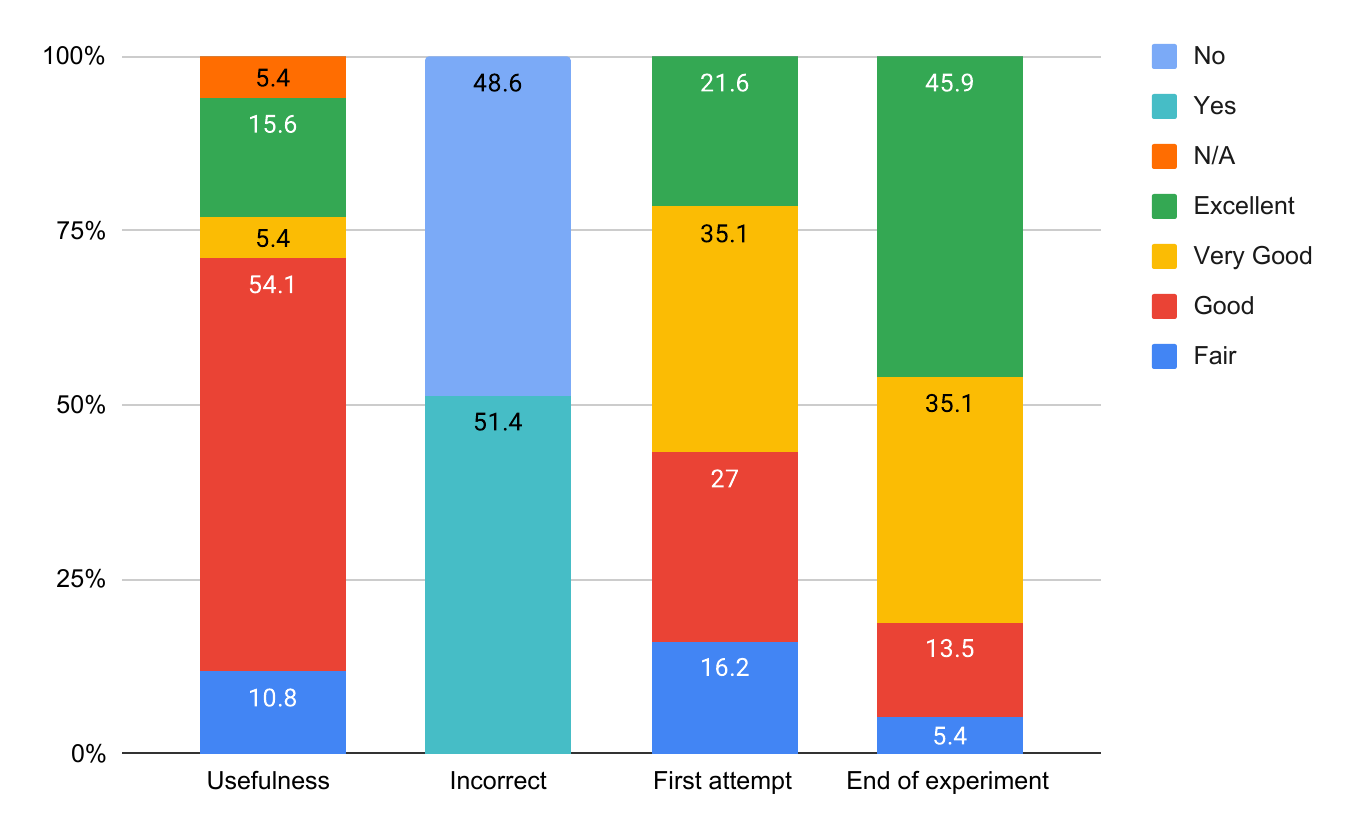}}
     \caption{RQ2: Qualitative results.}
    \label{fig:rq2_results}
\end{figure}

Figure \ref{fig:rq2_results} shows how the participants qualitatively assessed the responses of \emph{GPT-4}.
The first bar shows participants' prior perception of \emph{GPT-4}'s general usefulness.
More than 83.8\% of the participants were satisfied ($>=$ Good) with \emph{GPT-4}.
Some of the participants (5.4\%) have never used \emph{GPT-4} before.
The second bar shows if they encountered any incorrect responses on any steps during the study.
We could see that almost half (48.6\%) of the participants did not encounter any incorrect responses during the study.
The third bar shows their evaluation of the responses from the first attempt of prompts, and the last bar shows their evaluation at the end of the experiment.
The evaluations for initial responses were already considerably good with more than 83.8\% satisfactory score ($>=$ Good).
After applying the conversational prompt, the responses received higher evaluations with a 94.6\% satisfactory score ($>=$ Good), an increase of 10.8\% points.

We can see that their evaluations change positively from the initial responses to the end of the experiment (excellent +24.3\%), indicating that their conversational prompt helped \emph{GPT-4} in generating better responses.
This result is also consistent with previous studies that multiple conversational prompting improves the results of LLMs \cite{wu2022promptchainer,wu2022ai,trautmann2023large}.

\mybox{\textbf{Answer to RQ2}: Our qualitative analysis shows that participants were generally satisfied with \emph{GPT-4}'s responses. Satisfaction increased from 83.8\% to 94.6\% after using conversational prompts, a 10.8\% points improvement, indicating the benefit of conversational prompts.}

\subsection{RQ3: Trend in Forming Conversational Prompts}
\label{sec:rq3_result}

From analyzing the chat logs, we identified nine types of conversational prompts: (1) Request improvements with certain keywords, (2) Provide more context, (3) Add specific instructions, (4) Point mistakes then request fixes, (5) Ask questions to guide the correct way, (6) Request verification, (7) Request more examples, (8) Request more detailed description, and (9) Request another or a different version of generation.

Table \ref{tab:rq3_results} shows the categorized conversational prompts and their numbers used for each task.
We can see that the trends tend to differ per task.
For code summarization, we can see that requesting certain improvements with different keywords was the most common.
The keywords for requesting improvement are e.g., ``more natural'', ``abstract'', ``semantic-focused'', ``human-readable'', etc.
We suspect they requested these improvements due to \emph{GPT-4}'s verbose generations, which generate line-by-line explanations of the code's syntax. Code summarization focuses on automating the technical documentation process as it is a nontrivial task \cite{hu2020deep}. Explaining a method's syntax line-by-line would defeat the whole purpose of the task as developers would not read all the syntax explanations. To get the implicit summary of the method's functionality, we observe that most of the participants request improvements in abstracting and focusing on the semantics. Previous literature also points to the verbosity of LLMs' explanations of code \cite{macneil2023experiences, kabir2023answers}. In addition, some participants requested more descriptions for core logic, where syntax-level explanations are lacking, and added specific instructions to move the line-level explanations to the top of the method to resemble technical documentation, e.g., JavaDoc.
Since the output of the code summarization task is a natural language description of code, participants also request the model to generate descriptions with a certain style or format.

Adding specific instructions was found to be the most common for code generation tasks.
The instructions have a variety of ranges, e.g., ``remove the main method'', ``remove the import statements'', ``use a certain type for the variables/parameters'', ``remove external packages'', etc.
Close runner-ups were ``requesting improvements'' and ``asking questions to guide the model''.
The keywords used for requesting improvement in code generation are e.g., ``good format'', ``readable'', ``better quality'', ``most succinct working'', ``clear'', etc.
An example for asking questions to the model would be, e.g., ``Is the following code snippet syntactically correct?'', ``Is the generated code executable?'', ``Is this the most efficient version of code?'', etc.
These questions aren't the type to ask to perform a specific instruction but to question the reasoning path of the model to re-think their decision and then make changes if they were incorrect.
We suspect these conversational prompts are made as \emph{GPT-4} fails to infer some missing specifications that aren't explained in the abstract text description of code. To fill these gaps, participants would 1) add specific instructions, 2) ask for improvement of code quality, or 3) ask questions for verification or to guide the model into a different reasoning path. Generally, our findings are aligned with previous literature \cite{alomar2024refactor} in that developers tend to add specific instructions for refactoring (i.e., add/remove certain code elements) or request improvement code quality from LLMs. ``Asking questions for verification to guide the model to correct reasoning paths'' is a new pattern we found in this study. We also observed that participants requested to remove irrelevant generations (e.g., important statements, main method, external packages) due to LLM hallucination.

For the code translation task, the most common conversational prompt was to give certain instructions to perform.
These instructions include, e.g., ``using a certain type'', ``using a certain logic'', ``modifying the code to support certain data types'', ``removing import statements'', ``handling certain parameters'', etc.
The runner-up asked certain questions to verify or double-check if the generated code met the specific syntax of the target language.
Generally, the results were similar to code generation as their output is both source code.
Other commonly used conversational prompts include ``adding more context information'', ``requesting more details or descriptions'', ``pointing out mistakes then requesting fixes'', etc.
A previous study found that the most prevalent translation bug was data-related, i.e., data types, parsing input data, and output formatting issues \cite{pan2024lost}. We can observe that the participants try to address these problems by adding specific instructions like ``use a certain type'', ``modify to support certain data type'', or ``handle certain parameters''. We also observe that similar to the code generation task, they tried to remove irrelevant generations due to LLM hallucination. Previous studies have also reported that more than one-third of translation bugs are due to syntactic and semantic differences between languages. This finding is also in line with our previous findings in Section \ref{sec:task_specific} that they fail to generate certain keywords specific to the target language. We observe that participants tried to mitigate this by adding specific instructions or asking questions for target language-specific syntax.

\mybox{\textbf{Answer to RQ3:} Different trends were observed in conversational prompts per task. ``Requesting improvements with general keywords'' was most common for code summarization, while ``adding specific instructions'' was most frequent for code generation and translation. Participants used conversational prompts to address known LLM limitations, such as verbosity, missing information, and hallucination.}

\begin{table}[t!]
\centering
\caption{Trend: the number of different prompt categories participants used for each task.}
\label{tab:rq3_results}
\begin{adjustbox}{width=\linewidth}
\begin{tabular}{|l|c|c|c|}
\hline
\textbf{Prompts} & \textbf{ComGen} & \textbf{Codgen} & \textbf{CodTrans} \\ \hline
Request improvements & 12 & 7 & 5 \\ \hline
Request more description & 7 & - & - \\ \hline
Add specific instructions & 7 & 8 & 9 \\ \hline
Ask questions to find correct way & 6 & 7 & 7 \\ \hline
Add more context & 5 & 6 & 3 \\ \hline
Request examples & 3 & 1 & 1 \\ \hline
Request verification & 2 & 1 & 1 \\ \hline
Point mistake then request fix & 2 & 5 & 4 \\ \hline
Request another generation & 1 & - & 2 \\ \hline
\end{tabular}
\end{adjustbox}
\end{table}

\subsection{RQ4: Efficacy of Conversational Prompts}
\label{sec:rq4_result}
RQ4 aims to investigate which category of conversation prompts contributes to better results. 
Table \ref{tab:rq4_results} shows the top five ranks of conversational prompts that contribute the most to better results. 
``Requesting improvement'' was shown to be the most effective conversational prompt category for code summarization.
``Adding more context information'' was shown to improve code generation the most.
For code translation, ``asking questions to find the correct reasoning path'' was shown to have the best performance.
We can observe that the ranking of trends and the ranks that best improve the responses are very similar but not the same.
For example, the first rank in trend and improvement for code summarization were both requesting certain improvements.
For code generation, adding specific instruction was first ranked on the trend ranking but appeared second in the improvement ranking.
Adding more context was found to be the first for improvement but found to be the third most common in the trends.
For code translation, asking questions to find the correct path was the first rank in improvement but appears on the second in the trend ranking.
Adding specific instructions was the most common prompt but was found to be the third most effective in improvement.
From the observation, we can state that the conversational prompt that requested certain improvements was most effective in mitigating verbose and syntax-focused code summarization. For example,  ``please make summaries more natural and abstract'', ``improve this by generating semantic-focused and abstract summaries'', etc. The prompt resulted in \emph{GPT-4} generate that are closer to the ground truth, which are short phrases that summarize the functionality of the method.
For code generation, adding more context was most effective to mitigate the limitation of \emph{GPT-4} in inferring the missing steps that the abstract text description encapsulates. For example, they would add the class information, e.g., class signature, description of what the class does, and other public method signatures within the class. The model was able to infer the missing steps by the project- or class-specific knowledge provided by the additional context.  
For code translation, asking questions to find the correct way was the most effective in generating the correct syntax of different languages and data types. For example, ``what about the @Parameter annotation?'', ``The JSON.parseArray function needs to know the generic type T to know the format of the return value, how should I modify the code?'', etc. As the examples show, they would ask questions to guide the model in resolving the data issue.

The similarity of the two ranks can be interpreted as participants having a good knowledge of how to guide the models for better results, even though the ground truth wasn't provided.
It is interesting to find that ``requesting improvement'' actually helps generate better responses even though these prompts do not necessarily provide specific knowledge on how to improve.

We also report the average BLEU score from the responses of conversational prompts in Table \ref{tab:conversational}.
We can see that the average BLEU score of all conversational prompts is comparable to the two automated prompts, i.e., basic and task-specific prompts.
However, only the code translation task outperforms the better-automated prompt, i.e., task-specific prompt, by a mere 0.27\% points.
The reason is due to the large variance caused by the different types of conversational prompts.
The conversational prompt that produces the highest BLEU score outperforms the automated prompts by a big margin, 15.8\% points for code summarization, 18.3\% points for code generation, and 16.1\% points for code translation tasks.

\mybox{\textbf{Answer to RQ4:}  Overall, conversational prompts can improve code tasks. ``Requesting improvement'' worked best for code summarization, ``adding context'' for code generation, and ``asking questions'' for code translation. The rank of trends and improvements showed similar patterns, indicating participants effectively guided the models. The best conversational prompts significantly outperformed automated prompts, improving by 15.8\% points, 18.3\% points, and 16.1\% points for code summarization, generation, and translation, respectively.} 

\begin{table}[t!]
\centering
\caption{Improvement: Top-5 prompt categories that effectively improve the basic prompts.}
\label{tab:rq4_results}
\begin{adjustbox}{width=\linewidth}
\begin{tabular}{|l|l|l|l|}
\hline
\textbf{Rank} & \textbf{Code summarization} & \textbf{Code generation} & \textbf{Code translation} \\ \hline
1 & Request improvements. & Add more context. & Ask questions. \\ \hline
2 & Add more context. & Add instructions. & Point mistake then fix. \\ \hline
3 & Ask questions. & Request improvements. & Add instructions. \\ \hline
4 & Add instructions. & Ask questions. & Request improvements. \\ \hline
5 & Request verification. & Point mistake then fix. & Add more context. \\ \hline
\end{tabular}
\end{adjustbox}
\end{table}
\begin{table}[t!]
\centering
\caption{Result of conversational prompts (conv.) vs automated prompts. Bold indicates the highest score.}
\label{tab:conversational}
\begin{adjustbox}{width=0.9\linewidth}
\begin{tabular}{c|cccc|}
\cline{2-5}
 & \multicolumn{4}{c|}{\textbf{BLEU}} \\ \hline
\multicolumn{1}{|c|}{\textbf{Task}} & \multicolumn{1}{c|}{\textbf{Average conv.}} & \multicolumn{1}{c|}{\textbf{Highest conv.}} & \multicolumn{1}{c|}{\textbf{Basic}} & \textbf{Task-specific} \\ \hline
\multicolumn{1}{|c|}{CodSum} & \multicolumn{1}{c|}{36.22} & \multicolumn{1}{c|}{\textbf{55.75}} & \multicolumn{1}{c|}{35.76} & 39.95 \\ \hline
\multicolumn{1}{|c|}{CodGen} & \multicolumn{1}{c|}{53.15} & \multicolumn{1}{c|}{\textbf{69.16}} & \multicolumn{1}{c|}{57.33} & 50.85 \\ \hline
\multicolumn{1}{|c|}{CodTran} & \multicolumn{1}{c|}{67.43} & \multicolumn{1}{c|}{\textbf{83.26}} & \multicolumn{1}{c|}{52.10} & 67.16 \\ \hline
\end{tabular}
\end{adjustbox}
\end{table}


\subsection{Discussion}
In this section, we discuss the main findings of this study in terms of prompt engineering and summarize them as recommendations for future work.
Then, we discuss the trade-offs between prompt engineering and fine-tuning.

As the quantitative results indicate, even though \emph{GPT-4} is a larger model than its baselines, it does not always yield better results.
Even typical prompt engineering strategies, such as in-context learning, did not show significant improvement.
On the other hand, the qualitative user study revealed that the key potential of \emph{GPT-4} is on conversation-based prompting, which includes human feedback in the loop.

There are two takeaway messages from our results: (a) To leverage LLMs to their best, a sequence of back-and-forth queries with the model may be needed. Iterative processes such as search-based optimization or reinforcement learning-based agents might be useful to ``engineer'' prompts for ASE tasks with full automation. (b) Human feedback still plays a crucial role in optimizing the prompts, as shown by conversational prompting. 
Thus, to remove humans from the loop completely, future studies are needed to analyze developer-written prompts more carefully and extract patterns that can be implemented as rules, fitness functions, rewards, and policies.

To summarize our findings, we conclude that, like most engineering problems, the choice between prompt engineering and fine-tuning LLMs is a trade-off. The following are some of the main factors that play a role in the trade-off:
\begin{itemize}
    \item \textbf{Performance:} The most explicit aspect is the performance of the models. From the results, neither approach always dominates the other. However, conversational prompt engineering has shown to have great potential. 
    \item \textbf{Cost:} Although fine-tuning is usually considered more expensive than prompt engineering, depending on the subscription models of the commercial LLMs, inference fees can quickly become a huge burden. For example, running \emph{GPT-4} for this study costs around 1,500 USD. Therefore, developers who have a large number of instances for inference should consider using a cheaper model or even fine-tuning a smaller in-house model. 
    \item \textbf{Ease of Use:} Prompt engineering has a significant benefit from this aspect, especially when using natural language-based interfaces. It becomes more significant as we see participants with little to no background in LLMs can perform ASE tasks with \emph{ChatGPT-4}. In contrast, they would not be able to perform with fine-tuned models without a certain knowledge of deployment, configuration, and developing fine-tuned models.
    \item \textbf{Control:} Finally, another aspect of the trade-off is the level of control on the internal configurations and outputs. For example, with \emph{GPT-4}, although you can set the temperature to zero, you still get non-deterministic answers, which makes the verification of the models more difficult and thus less interesting to some users.  
\end{itemize}

\section{Threats to Validity}
\label{sec:threats}

\textbf{Internal Validity:} A key threat is whether the designed prompts truly reflect the intended strategy or if better approaches could exist. For example, poor results of \emph{GPT-4} with in-context learning might be due to confounding factors, such as suboptimal sample choices or vocabulary. To address this, we reported the exact prompts used and conducted a qualitative survey where participants created their own prompts, providing a broader view of prompt engineering effectiveness.
{Another potential threat to validity arises from not explicitly examining biases such as task order and participant learning effects, though we mitigated order-related risks by randomizing task sequences across participants. The implications of participant segregation, while a compelling avenue for future analysis, were beyond the scope of this work due to brevity and will be explored in future research.}

\textbf{Construct Validity:} We used common metrics, such as BLEU scores, as in the original benchmarks. However, these metrics may not fully capture what they aim to measure. In the qualitative study, participants' evaluations of \emph{GPT-4}'s responses might be subjective. To mitigate this, we also performed a manual analysis by three co-authors. Investigating better human-aligned metrics remains an open area in the field.

\textbf{Conclusion Validity:} The non-deterministic behavior of LLMs \cite{ouyang2023llm} means results may not be reproducible. We set the temperature parameter to zero to increase determinism but acknowledge this does not eliminate the issue entirely. We provide generated responses from the quantitative study and raw chat logs from the survey to address this.
{A further limitation lies in the absence of statistical tests for quantitative comparisons, as fine-tuned model baseline distributions were unavailable (benchmark scores were sourced directly from leaderboards), and the qualitative analysis's limited sample size (6 samples per task) precluded meaningful statistical testing. While these constraints temper the robustness of our conclusions, we intend to address them in future work through expanded datasets and formal statistical evaluations.}

\textbf{External Validity:} We used recent state-of-the-art fine-tuned models for each task, but new baselines could alter our observations. Although we used widely accepted benchmark datasets, they do not represent all domains. For the survey, we used 18 samples to limit participant time, acknowledging that this limits generalizability. We chose a 1-hour survey duration to recruit more participants, balancing practical constraints like participant availability and budget. To mitigate this threat, we considered diverse domains and difficulty levels in the examples. We also included three different project types (library, algorithm, and web) to cover different domains. While in-context learning was one of the best automated prompting strategies at the start of our study, newer methods like Retrieval Augmented Generation (RAG) \cite{chen2024benchmarking} and Self-Reflection \cite{shinn2024reflexion} could affect our findings, warranting further study.

\section{Conclusion}
\label{sec:conclusion}
In this paper, we evaluated \emph{GPT-4} using basic, in-context learning, and task-specific prompts and compared them with fine-tuned language models. We also conducted a user study with 27 academic and 10 industry participants to assess the quality of \emph{GPT-4}'s responses and examine how participants designed conversational prompts. Our quantitative results show that \emph{GPT-4} does not significantly outperform the baselines. Qualitative results indicate that participants often request improvements, add context, or give specific instructions, which helps \emph{GPT-4} generate better responses. Our study suggests that \emph{GPT-4} has great potential for code tasks but requires careful human verification and interpretation.

\bibliographystyle{IEEEtran}
\bibliography{paper}
\balance
\end{document}